\documentclass[aps,prl,twocolumn,showpacs]{revtex4}
\usepackage{graphicx}

\begin{document}

\title{Magnetic scattering of spin polarized carriers in (In,Mn)Sb dilute magnetic semiconductor}

\author{M.~Csontos$^{1}$, T.~Wojtowicz$^{2,3}$,
X.~Liu$^{2}$, M.~Dobrowolska$^{2}$ B.~Jank\'o$^{2}$,
J.~K.~Furdyna$^{2}$ and G.~Mih\'aly$^{1}$}

\affiliation{$^{1}$Department of Physics, Budapest University of
Technology and Economics and \\
"Electron Transport in Solids" Research Group of the Hungarian
Academy of Sciences, 1111 Budapest, Hungary\\
$^{2}$Department of Physics, University of Notre Dame, Notre Dame,
Indiana 46556, USA\\
$^{3}$Institute of Physics, Polish Academy of Sciences, 02-668
Warsaw, Poland}

\date{\today}

\begin{abstract}

Magnetoresistance measurements on the magnetic semiconductor
(In,Mn)Sb suggest that magnetic scattering in this material is
dominated by isolated Mn$^{2+}$ ions located outside the
ferromagnetically-ordered regions when the system is below
$T_{c}$. A model is proposed, based on the $p$-$d$ exchange
between spin-polarized charge carriers and localized Mn$^{2+}$
ions, which accounts for the observed behavior both below and
above the ferromagnetic phase transition. The suggested picture is
further verified by high-pressure experiments, in which the degree
of magnetic interaction can be varied in a controlled way.

\end{abstract}


\maketitle

The presence of charge and spin degrees of freedom in III-Mn-V
magnetic semiconductors alloys opens new perspectives for
spintronic applications \cite{Ohno2000,Chiba2003,Yamanouchi2004}.
In these systems the Mn$^{2+}$ ions provide $S$=5/2 magnetic
moments, and also act as a source of valence band holes that
mediate Mn$^{2+}$-Mn$^{2+}$ interactions. The interactions between
the randomly positioned magnetic ions are predominantly
ferromagnetic (FM), resulting in a low temperature FM phase
\cite{Ohno1998,Dietl2000,Dietl2001,Konig2002,Dietl2002,Jungwirth2002}.
Thus the magnetic and the electrical transport properties of
III-Mn-V materials are fundamentally inter-dependent. Although
many aspects of free carrier magnetotransport have already been
explored in these materials, its microscopic understanding is
still far from complete. Recent scaling theory on
Anderson-localized disordered ferromagnets applies to transport in
the \emph{localized} limit \cite{Zarand2005}, but the more
interesting "metallic" systems - where the mean free path
satisfies the $k_{\rm F}l>>$ 1 condition (where $k_{\rm F}$ is the
Fermi wave number) - require a different approach.

Many spintronic applications envision spin injection and/or
spin-polarized current, and are thus expected to hinge critically
on our understanding of spin scattering of the charge carriers. In
this paper we present a detailed experimental study of magnetic
scattering processes in the dilute magnetic semiconductor
(In,Mn)Sb. We find that in the FM phase the positional disorder of
the magnetic moments is reflected in the slowly saturating nature
of both the magnetization and magnetoresistance. This suggests the
presence of isolated Mn$^{2+}$ ions located outside the FM-ordered
regions. The large negative magnetoresistance observed below
$T_{c}$ is attributed to a first order magnetic scattering of
spin-polarized holes on such isolated paramagnetic Mn$^{2+}$ ions.
Above $T_{c}$ the spontaneous spin polarization of the carriers is
lost and second-order spin-independent processes begin to
dominate. This picture is supported by high pressure measurements,
in which the interaction between charge-carrying holes and
localized Mn$^{2+}$ ions can be experimentally varied
\cite{Csontos2005}.

In the present work we present results obtained on
In$_{1-x}$Mn$_{x}$Sb for $x$ = 0.02. Similar behavior has been
found for other concentrations. More importantly the same
\emph{qualitative} behavior has also been reported for the most
widely studied magnetic semiconductor Ga$_{1-x}$Mn$_{x}$As
\cite{Ohno1998,Iye1999}. We therefore believe that the results on
(In,Mn)Sb reflect a common feature of highly-conducting III-Mn-V
magnetic semiconductors. While specific details of the spin
landscape (FM-ordered regions and isolated Mn$^{2+}$ spins) may
vary from system to system, the great advantage of
In$_{0.98}$Mn$_{0.02}$Sb is that here the isolated Mn$^{2+}$ ions
can be unambiguously identified as the dominant source of the
magnetic scattering.

\begin{figure}[t!]
\includegraphics[width=\columnwidth]{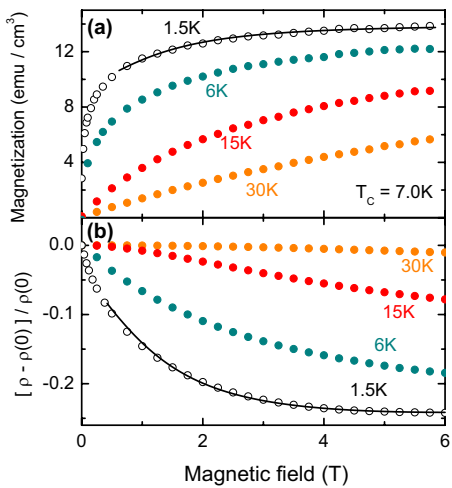}
\caption{(Color online) Magnetization measured by magnetic
circular dichroism (a) and reflected in magnetoresistance (b) as a
function of the applied field up to 6T.  The Curie temperature is
$T_{c}$=7K. The solid lines represent the B$_{5/2}$ Brillouin
function characterizing the magnetization of isolated Mn$^{2+}$
ions.} \label{intro.fig}
\end{figure}

The In$_{1-x}$Mn$_{x}$Sb films were grown by low-temperature
molecular beam epitaxy (MBE) in a Riber 32 R\&D MBE system on
closely lattice-matched hybrid (001) CdTe/GaAs substrates to a
thickness of about 230 nm \cite{Wojtowicz2003}. (For further
growth details and structural characterization see Ref. 13).  In
the In$_{0.98}$Mn$_{0.02}$Sb sample used in this study 80\% of the
Mn is magnetically active while the compensation by interstitial
Mn is 5\% \cite{Wojtowicz2003}. These numbers are consistent both
with the estimated saturation magnetization of about 13
emu/cm$^{3}$ (see Fig.~\ref{intro.fig}) and with the hole
concentration of $\approx$ 3$\cdot$10$^{-20}$ cm$^{-3}$ deduced
from Hall effect measurements. At low temperatures the resistivity
is typically 250 $\mu\Omega$cm, corresponding to a metallic mean
free path of $k_{\rm F}l\approx$ 30.

The magnetic properties were investigated by magnetic circular
dichroism (MCD) and by superconducting quantum interference device
(SQUID) measurements. The MCD signal was normalized to low-field
SQUID data, with an accuracy of about 20\%. Magnetotransport
measurements were carried out in a six-probe arrangement, with
magnetic field normal to the layer. For high pressure measurements
the samples were mounted in a self-clamping cell, with kerosene as
the pressure medium.

The field dependence of magnetization $M(B)$ obtained from MCD
measurements is shown in the upper panel of Fig.~\ref{intro.fig}.
At T = 30K (well above the Curie-temperature) $M(B)$ is seen to
vary linearly with the applied field, as expected for a
paramagnetic material. Below $T_{c}$ = 7K the almost step-like
$M(B)$ curves resemble the magnetization of a FM material, but no
saturation is reached up to a few Tesla. For example, at $T$ = 1.5
K about 80\% of the estimated saturation value is achieved at $B$
= 1 T, and we relate this large value of $M$ to the spin alignment
within the FM-ordered regions of the system. On the other hand,
the slowly-saturating part superimposed on this FM magnetization
clearly indicates that a fraction of the Mn$^{2+}$ magnetic
moments still remain isolated (i.e., are not included in the
FM-ordered regions).

\begin{figure}[b!]
\includegraphics[width=\columnwidth]{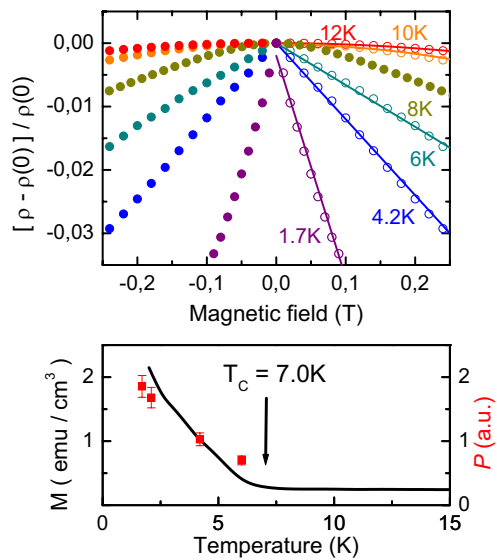}
\caption{(Color online) (a) Low-field magnetoresistance, showing
quadratic field dependence above $T_{c}$, and linear dependence in
the FM phase. (b) Remanent magnetization measured by SQUID as a
function of temperature (solid line). The data points show the
spin polarization of charge carriers deduced from the initial
slope of the magnetoresistance below $T_{c}$ (see text).}
\label{lowfield.fig}
\end{figure}

\begin{figure}[t!]
\includegraphics[width=\columnwidth]{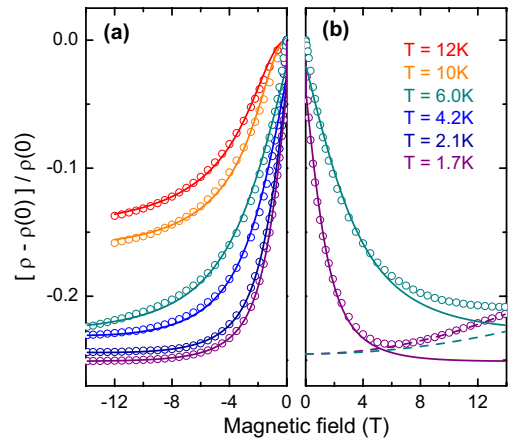}
\caption{(Color online) Magnetoresistance measured up to 14 T at
selected temperatures below (various blue curves) and above
$T_{c}$ (red and yellow curves). A positive quadratic term
attributed to normal magnetoresistance is subtracted at each
temperature, as shown in the right panel. Continuous lines in the
left panel correspond to Eqs.~\ref{ferromagnetic} and
\ref{paramagnetic}, with $\left\langle S_{z}\right\rangle$ given
by the Brillouin function for $S$=5/2. The only fitting parameter
is the correlation term $T_{\rm eff}$ in the argument $\alpha$ of
the $B_{5/2}(\alpha)$ function. ($\alpha=g\mu_{B}B/k_{B}T_{\rm
eff}$, where $T_{\rm eff}$ below and above $T_{c}$ is,
respectively, $T+T_{\rm AF}$ and $T-T^{*}$.)} \label{temp.fig}
\end{figure}

\begin{figure}[t!]
\includegraphics[width=\columnwidth]{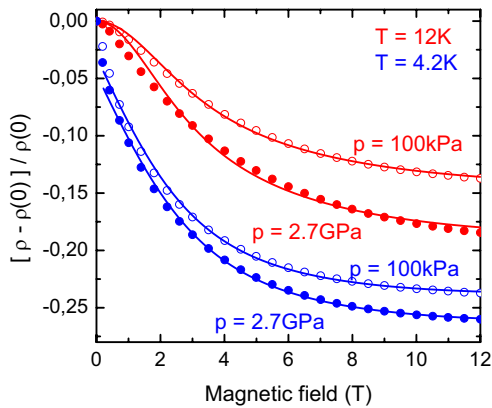}
\caption{(Color online) Effect of hydrostatic pressure on
magnetoresistance measured below (blue curve) and above $T_{c}$
(red curve). Open symbols: ambient pressure; solid symbols: $p$ =
2.7GPa. Fits for ambient pressure are calculated using
Eqs.~\ref{ferromagnetic} and \ref{paramagnetic}; fits for pressure
data are calculated using the same expression, with $J_{pd}$
increased by 5.6\%. In the paramagnetic phase we found that
$T^{*}$ is also enhanced by pressure (from 7.1 K to 7.6 K), as
expected.} \label{pressure.fig}
\end{figure}

The magnetoresistance curves measured at the same temperatures and
in the same field range are shown in the lower panel of
Fig.~\ref{intro.fig}. At high fields they exhibit a similar field
dependencies as the corresponding magnetization curves. The
simultaneous slow saturation observed in magnetization and in
resistivity suggests that the magnetoresistance arises from a
magnetic scattering process which is gradually "frozen out" as
more isolated random Mn$^{2+}$ spins become aligned. The analysis
given below for the results in
Figs.~\ref{lowfield.fig}-\ref{pressure.fig} shows that the field-
induced alignment indeed follows a $B_{5/2}$ Brillouin function,
clearly suggesting that the slow component of the saturation is
closely related to the isolated Mn$^{2+}$ spin population.

Figure~\ref{lowfield.fig} presents the magnetoresistance at low
fields. Note that the quadratic behavior of $\rho$ above the phase
transition changes to linear as $T$ is decreased below $T_{c}$. We
shall show that the \emph{linear} magnetoresistance observed in
the FM phase indicates the presence of spin-polarized carriers in
the scattering process. Specifically, such linear $B$-dependence
is \emph{symmetry-breaking}, and indicates that the polarization
of the charge-carrying holes changes sign as $B$ is reversed.

In the following we briefly discuss the expected behavior when
spin-polarized carriers are scattered on isolated Mn$^{2+}$ ions.
The Hamiltonian for the interaction between delocalized holes of
spin s and half-filled 3$d$ shells of the Mn$^{2+}$ ($S$=5/2) ions
is
\begin{equation}
H=V-2J_{pd}\mathbf{s}\cdot\mathbf{S}\mbox{ ,} \label{hamiltonian}
\end{equation}
where $V$ is the spin-independent part of the potential, and the
second term describes the exchange interaction described by the
$J_{pd}$ coupling constant \cite{Dietl2001,Beal-Monod1967}. The
spin-dependent scattering on isolated (paramagnetic) Mn$^{2+}$
ions can be described by a perturbation calculation in the limit
$|J_{pd}|\ll V$. In the presence of a magnetic field the
relaxation times $\tau_{\pm}$ for the spin-up and spin-down
carriers is (up to second order in $J_{pd}/V$) given by
\cite{Beal-Monod1967}
\begin{eqnarray}
\frac{1}{\tau_{\pm}}=\mathcal{C}\Bigg[1\mp\frac{2J_{pd}}{V}\left\langle
S_{z}\right\rangle+\left(\frac{2J_{pd}}{V}\right)^{2}f(\left\langle
S_{z}^{2}\right\rangle)\Bigg]\mbox{ ,} \label{bealmonod}
\end{eqnarray}
where $\left\langle S_{z}\right\rangle$ is the average
field-direction component given by the $B_{5/2}(\alpha)$ Brillouin
function with an argument of $\alpha=g\mu_{B}B/k_{B}T_{\rm eff}$,
$M = g\mu_{B}\left\langle S_{z}\right\rangle$ is the corresponding
magnetization, while $f(\left\langle S_{z}^{2}\right\rangle)$ and
$\mathcal{C}$ are given in Ref.~14.

In a system containing a few percent of Mn the $p$-$d$ interaction
defined in Eq.~\ref{hamiltonian} couples the Mn$^{2+}$ moments to
each other via an RKKY-like superexchange (for details see
Ref.~6). Since the carrier density is low, the first node of the
oscillating RKKY function occurs at a length scale larger than the
average Mn-Mn separation, giving rise to a net FM interaction,
$J\propto J_{pd}^{2}$. Magnetic ordering in such spin landscape is
somewhat unusual: according to Monte Carlo simulations
\cite{Bhatt2001} small FM-ordered regions in fact begin to develop
\emph{above} the mean-field $T_{c}$ where the local Mn density
exceeds the average. These regions increase as T decreases,
eventually developing into a percolated network at $T_{c}$.
However, a fraction of the spins can still remain outside this FM
network even below $T_{c}$, and we suggest that these constitute
the source of magnetic scattering.

In the FM phase the numbers of spin-up and spin-down carriers are
different, so that the leading linear term in $\left\langle
S_{z}\right\rangle$ does not cancel out in Eq.~\ref{bealmonod},
and dominates over the second-order quadratic term. The difference
between $1/\tau_{+}$ and $1/\tau_{-}$ then gives rise a
non-vanishing $2\mathcal{P} \cdot\frac{2J_{pd}}{V}\left\langle
S_{z}\right\rangle$ term, where $\mathcal{P}$ is the
spin-polarization of the charge carriers. Consequently, the first
order magnetic scattering leads to a magnetoresistance that is
proportional both to the spin polarization arising from the FM
block and to the magnetization of the isolated paramagnetic
moments:
\begin{equation}
\frac{\rho(B)-\rho(0)}{\rho(0)}=-4\mathcal{P}\frac{J_{pd}}{V}\left\langle
S_{z}\right\rangle = -4\mathcal{P}\frac{J_{pd}}{g\mu_{B}V}
M(B)\mbox{ .}\label{ferromagnetic}
\end{equation}
An important prediction of this Eq.~\ref{ferromagnetic} is that in
the FM phase the magnetoresistance directly measures the
magnetization of the isolated Mn$^{2+}$ ions. This is verified by
the high field results shown in Fig.~\ref{temp.fig}, where the
continuous lines correspond to the $B_{5/2}$ Brillouin function
describing the magnetization of isolated $S$=5/2 spins. The
argument of the Brillouin function is
$\alpha=g\mu_{B}B/k_{B}T_{\rm eff}$, where the effective
temperature $T_{\rm eff} = T+T_{\rm AF}$ contains an empirical
antiferromagnetic (AF) coupling parameter describing the natural
AF interaction between Mn ions \cite{Dietl2001,Shapira1986}.
$T_{\rm AF}$ is the \emph{only} fitting parameter used to describe
the observed \emph{magnetic field - temperature} scaling. The same
(small) value $T_{AF}$ = 1.4K has been used to fit all curves.

At low T a quadratic contribution to the magnetoresistance
gradually emerges, attributed to the ordinary positive
magnetoresistance. Although quite small, it is clearly visible at
1.7 K, where the Brillouin term becomes saturated at around
$B\approx$ 6T. This quadratic contribution is shown in the left
panel of Fig.~\ref{temp.fig} for various temperatures.

We believe that the $B_{5/2}$ Brillouin fits in
Fig.~\ref{temp.fig} for the FM phase provide strong evidence for
magnetic scattering of spin-polarized holes on isolated Mn$^{2+}$
($S$=5/2) ions. Eq.~\ref{ferromagnetic} also contains a
temperature-dependent spin polarization $\mathcal{P}$ of the
carriers, which can be directly deduced from the low field slope
of the magnetoresistance curves (Fig.~\ref{lowfield.fig}, upper
panel), as the initial variation is proportional to
$\mathcal{P}(T)\cdot g\mu_{B}B/k_{B}T_{\rm eff}$. Comparison of
the latter with remanent magnetization measured by SQUID is shown
in the lower panel of Fig.~\ref{lowfield.fig}.
$\mathcal{P}(T)\propto M_{\rm rem}(T)$ relation in the FM phase
provides additional support for the role of spin polarization in
our model.

Above $T_{c}$, i.e., in the paramagnetic phase, the global spin
polarization is lost, and for $\alpha <$ 2  (which is fulfilled
for our experiments) the second order term in Eq.~\ref{bealmonod}
leads to magnetoresistance expressed by \cite{Beal-Monod1967}:
\begin{eqnarray}
\frac{\rho(B)-\rho(0)}{\rho(0)}&=&
-\frac{J_{pd}^2}{V^2}\Bigg[4\left\langle S_{z}\right\rangle^2+ \nonumber \\
&+&\left\langle S_{z}\right\rangle\left(\coth\frac{\alpha}{2} -
\frac{\alpha}{2\sinh^2\frac{\alpha}{2}}\right)\Bigg]\mbox{ .}
\label{paramagnetic}
\end{eqnarray}
Equation~\ref{paramagnetic} accounts for the low-field quadratic
variation of magnetoresistance above $T_{c}$ seen in
Fig.~\ref{lowfield.fig}, as well as for its behavior over the
broad magnetic field range shown in Fig.~\ref{temp.fig}. Although
a second-order term is expected to make a weaker contribution, the
observed magnetoresistance is still very large, since in the
paramagnetic phase most of the Mn ions take part in the
scattering. In this picture we neglect fluctuations in the Mn
concentration, and the ferromagnetic interaction between the Mn
ions in the paramagnetic phase is expressed via an effective
temperature $T_{\rm eff}$ for that phase,  $T_{\rm eff}=T-T^{*}$,
where simple mean field theory predicts $T^{*}$ close to $T_{c}$.
Note that $T_{\rm eff}$ is the only fitting parameter used in the
calculations. The fits shown in Fig.~\ref{temp.fig} indeed yield
$T^{*}$ very close to $T_{c}$ = 7.0K: ($T^{*}$ = 6.2 and $T^{*}$ =
7.1 K for the orange and the red curves, respectively), thus
providing support for the intuitive mean-field scenario.
The validity of Eqs.~\ref{ferromagnetic} and \ref{paramagnetic}
can be further tested by applying hydrostatic pressure to tune the
exchange coupling $J_{pd}$. Earlier investigations on various
magnetic semiconductors \cite{Dietl2001,Giebultowicz1990}
indicated that the $p$-$d$ exchange energy varies inversely as the
volume $v_{0}$ of the unit cell, $J_{pd}\propto v_{0}^{-1}$. A
clear pressure-induced enhancement of magnetoresistance is shown
in Fig.~\ref{pressure.fig} both below and above $T_{c}$. Using the
bulk modulus of InSb ($\kappa$ = 48 GPa) one expects a $\delta
v_0/v_0=p/\kappa=5.6$ \% decrease in $v_{0}$ for the applied
pressure of $p$ = 2.7 GPa, and thus the same increase in $J_{pd}$.

In the FM phase magnetic scattering varies as the product of
$J_{pd}$ and $\mathcal{P}$, which itself varies as the strength of
the $J_{pd}$ coupling. Taking both effects into account,
Eq.~\ref{ferromagnetic} predicts that below $T_{c}$
magnetoresistance depends on the $p$-$d$ coupling as $J_{pd}^2$.
Accordingly, the application of $p$ = 2.7 Gpa should result in an
11.2\% enhancement of the magnetoresistance. Indeed, in
Fig.~\ref{pressure.fig} excellent agreement is obtained when we
fit the high pressure $T$ = 4.2 K data simply by applying this
11.2\% magnification to the ambient fit (the fitting parameter
$T_{\rm AF}$ is influenced only slightly: it increases by 0.38 K).

Above $T_{c}$ one should - besides the $J_{pd}^{2}$ enhancement -
also take into account the pressure-induced increase of FM
coupling described by the parameter $T^{*}$. Fits of high pressure
data at 12 K indicate that $T^{*}$ is indeed enhanced from 7.1 K
to 7.6 K, providing additional support to the consistency of our
picture.

In conclusion, the high-field and high-pressure transport
experiments in (In,Mn)Sb revealed that the large magnetoresistance
observed in this magnetic semiconductor is dominated by scattering
on isolated Mn$^{2+}$ ($S$=5/2)) ions. The qualitative changes
(e.g., the change from quadratic to linear magnetoresistance as
the temperature is lowered below $T_{c}$, or the freezing out of
scattering as spins become aligned by an applied field) can all be
described by the $p$-$d$ interaction between charge-carrying holes
and localized Mn moments. In the FM phase the magnetoresistance
arises from scattering of spin polarized carriers, its field
dependence directly following the magnetization of the few ions
that are not included in the FM order. Above $T_{c}$ the spin
polarization is lost and magnetic scattering becomes a second
order process, but now \emph{most} Mn ions participate in the
scattering, so that the magnitude of the effect remains large.
Finally, by enhancing the strength of the $p$-$d$ coupling by
hydrostatic pressure, we also showed that the observed
pressure-induced increase of magnetoresistance also follows
directly from the above microscopic picture.

Valuable discussions with T. Dietl and G. Zar\'and are
acknowledged. This research has been supported by the Hungarian
research funds OTKA TS049881, T037451; and the National Science
Foundation NIRT Award DMR 02-10519. B.~J. was also supported by
the Alfred P. Sloan Foundation.

\end{document}